\documentclass[sn-mathphys-num]{sn-jnl}

\usepackage{graphicx}%
\usepackage{multirow}%
\usepackage{amsmath,amssymb,amsfonts}%
\usepackage{amsthm}%
\usepackage{mathrsfs}%
\usepackage[title]{appendix}%
\usepackage{xcolor}%
\usepackage{textcomp}%
\usepackage{manyfoot}%
\usepackage{booktabs}%
\usepackage{algorithm}%
\usepackage{algorithmicx}%
\usepackage{algpseudocode}%
\usepackage{listings}%

\begin{document}

\title[Article Title]{Bootstrap Prediction and Confidence Bands for Frequency Response Functions in Posturography}


\author[1,2]{\fnm{Vittorio} \sur{Lippi}}\email{vittorio.lippi@uniklinik-freiburg.de}\email{vittorio\_lippi@hotmail.com}

%

\affil[1]{\orgdiv{Institute of Digitalization in medicine}, \orgname{Faculty of Medicine and Medical Center - University of Freiburg}, \orgaddress{\street{Breisacher Stra{\ss}e 153}, \city{Freiburg im Breisgau}, \postcode{79110}, \state{BW}, \country{Germany}}}
\affil[2]{\orgdiv{Department of Neurology and Neurophysiology}, \orgname{Faculty of Medicine and Medical Center - University of Freiburg}, \orgaddress{\street{Breisacher Stra{\ss}e 64}, \city{Freiburg im Breisgau}, \postcode{79110}, \state{BW}, \country{Germany}}}


\abstract{The frequency response function (FRF) is an established way to describe the outcome of experiments in posture control literature. The FRF is an empirical transfer function between an input stimulus and the induced body segment sway profile, represented as a vector of complex values associated with a vector of frequencies. For this reason, testing the components of the FRF independently with Bonferroni correction can result in a too-conservative approach. Performing statistics on scalar values defined on the FRF, e.g., comparing the averages, implies an arbitrary decision by the experimenter. This work proposes bootstrap prediction and confidence bands as general methods to evaluate the outcome of posture control experiments, overcoming the foretold limitations of previously used approaches.}




\maketitle

\section{Introduction}
\textbf{The frequency response function (FRF)} is a usual way to describe the outcome of experiments in posture control literature. In particular, the FRF is an empirical transfer function between an input stimulus and the induced body movement with an additional step of averaging the value at some sets of frequencies, as explained in detail in section \ref{FRF}. By definition, the FRF is a complex function of frequency. When statistical analysis is performed to assess differences between groups of FRFs (e.g., obtained under different conditions or from a group of patients and a control group), the FRF's structure should be considered. Usually, the statistics are performed by defining a scalar variable to be studied, such as the norm of the difference between FRFs, or considering the components independently that can be applied to real and complex components separately\cite{lippi2023human,Akcay2021}, in some cases both approaches are integrated, e.g., the comparison frequency-by-frequency is used as a post-hoc test when the null hypothesis is rejected on the scalar value\cite{lippi2020body}. The two components of the complex values can be tested with multivariate methods such as Hotelling's T2 as done in \cite{asslander2014sensory} on the averages of the FRF over all the frequencies, where a further post hoc test is performed applying bootstrap on magnitude and phase separately. The problem with the definition of a scalar variable as the norm of the differences or the difference of the averages in the previous examples is that it introduces an arbitrary metric that, although reasonable, has no substantial connection with the experiment unless the scalar value is assumed a priori as the object of the study as in \cite{lippi2020human} where a human-likeness score for humanoid robots is defined on the basis of FRFs difference. On the other hand, testing frequencies (and components) separately does not consider that the FRF's values are not independent and applying corrections for multiple comparisons, e.g., Bonferroni, can result in a too-conservative approach destroying the power of the experiment. In order to properly consider the nature of the FRF, a method oriented to complex functions should be used. In \cite{Lippi2023} a preliminary method based on random field theory inspired by \cite{pataky2016region} was presented: to take into account the two components (imaginary and real) as two independent variables, the fact that the same subject repeated the test in the two conditions, a 1-D implementation of the Hotelling T2 is used as presented in\cite{pataky2014vector} but applied in the frequency domain instead of the time domain.

Several approaches to statistical analysis of continuous data based on confidence bands are reviewed in \cite{joch2019inference}. In particular, the \textit{function based resampling technique} (FBRT) is based on fitting data into mathematical functions and using its coefficients to estimate a variability function. A scaling factor is used to adjust the prediction coverage. The functions used are usually sinusoid; hence, the parameters are the coefficients of Fourier expansions. The limitation of FBRT is that the choice of the functions (i.e., of the frequencies) is not always obvious: while in human movement signals such as joint trajectories during gait assuming a \textit{low-pass} power-spectrum for the samples makes \textcolor{black}{it} reasonable to truncate its Fourier expansion, i.e. take the first $N$ terms \cite{lenhoff1999bootstrap}, other biological signals like EEG may not clearly exhibit a set of frequencies that are representative in all the samples \cite{joch2019inference}. In the specific case of FRFs from posture control experiments, the sample itself, by definition, provides a predefined set of frequencies (as it will be clear in \S \ref{FRF}). This suggests that all the FRFs in the samples can be conveniently transformed to time domain signals that consistently represent all the components of the FRFs. On this basis, the present work proposes a bootstrap method for computing prediction and confidence bands. The method is implemented by transforming the FRF in the time domain, obtaining a \textit{pseudo-impulse-response} (PIR) function on which such bands can be defined and computed. The results of the tests in the frequency domain are then presented and discussed.

\textbf{Prediction band} represents the \textcolor{black}{region} that contains with a given probability $\alpha\%$ a new PIR drawn from the same population of the sample under analysis. This can be used to classify a new observation as belonging or not to the sample population on which the band is computed.

\textbf{Confidence band} for the population's mean \textcolor{black}{PIR} can also be defined with a probability $\alpha\%$. It can be used to test the effect of a condition or the difference between groups, e.g., the null hypothesis for an experiment is that the mean difference between the PPR under two conditions is zero.

\textbf{This paper} proposes the bootstrap method to compute the \textcolor{black}{regions} mentioned above following the one proposed for trajectories in \cite{lenhoff1999bootstrap} but adapted to the specific case of FRF. In the \textit{Methods} section, a definition of the FRF used in posture control literature is provided for reference, and then the Bootstrap method is described. Then, a way to visualize the \textcolor{black}{regions} in the complex space is proposed, and finally, tests to evaluate the method are presented, using real human data from posture control experiments. The Results section presents the results of the tests. In the \textit{Discussion}, the results are analyzed with emphasis on the implications for posturography and potentially for evaluating humanoid robots' posture control. In the last section, \textit{The Code}, there are links to the repository to download the Matlab \cite{MATLAB:2019b} source and instructions to run the functions computing the bands and running tests.

\section{Methods}
\subsection{The FRF}
\label{FRF}
\begin{figure}
\centering
\includegraphics[width=0.90\textwidth]{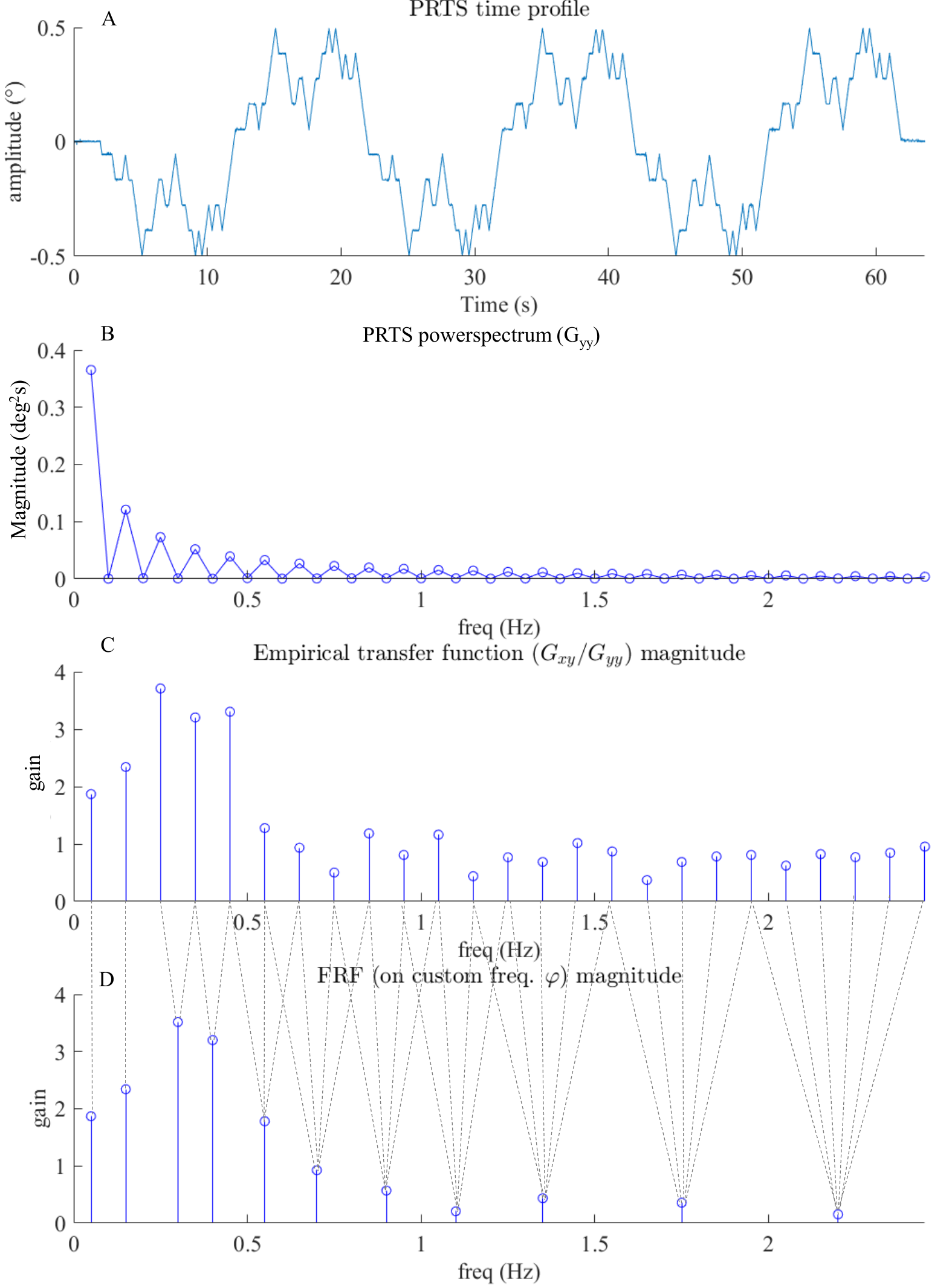}
\caption{An example of FRF. A: Time profile of the PRTS signal. B: Magnitude of the PRTS DFT, notice the \textit{comb} structure with peaks alternated to zeroes. C: magnitude of a sample empirical transfer function is defined {at} the frequencies associated with non-zero values in the input PRTS. D: FRF is obtained by averaging the transfer function over different ranges of frequencies. In the final FRF, the average over a range of frequencies is associated with the mean of the frequencies in such range. The dotted lines connect the values of the FRF with the samples averaged to obtain them.Although the plots show just the magnitude for ease of reading, the average is computed in a complex domain, and the FRF is a complex function.}
\label{fig:FRF}
\end{figure}

The frequency response function, FRF, is an empirically computed transfer function defined between an input and output. In posture control experiments, the input is a stimulus, such as a support surface tilt with a specific profile, and the output is the body sway or the sway of a body segment. In this work, the FRF is defined as proposed in \cite{peterka2002sensorimotor}, in detail:
\begin{enumerate}
\item The input has a pseudo-random ternary signal profile, PRTS \cite{davies1970system}, composed of segments at constant speed $(0,+s,-s)$ where $s$ is a fixed speed value usually set to get a specific amplitude of the position profile (e.g., $1^\circ$ peak-to-peak tilt). {A PRTS profile is shown in Fig \ref{fig:FRF}A. Different PRTS signals can be generated with different seeds and settings. This particular profile was used in \cite{peterka2002sensorimotor} and proposed in several subsequent works.} 
\item The PRTS has a power spectrum characterized by peaks separated by zones with zero power, as shown in Fig. \ref{fig:FRF}B. The response is computed on such frequencies: Sway responses are averaged across all PRTS sequence repetitions across subjects, discarding the first cycle of each trial to avoid transient response effects. Spectra of the stimulus and body sway in space are computed using the Fourier transform. Finally, frequency response functions are computed as cross-power spectrum $G_{xy}(f)$ divided by the stimulus power spectra $G_{yy}(f)$ that is $G_{xy}/G_{yy}$.
\item In the examples presented in this paper, the signals are sampled at 50 Hz.
\item The FRFs are obtained from human experiments with healthy subjects described in \cite{icinco23,lippi2023human,lippi2020human,robovis21}
\item The values of the obtained transfer function are averaged over bands of frequencies as shown in Fig. \ref{fig:FRF}D, with the resulting FRF being represented by a vector of 11 complex values\footnote{in the original formulation \cite{peterka2002sensorimotor} the frequencies were 22 corresponding to a stimulus that was twice long, i.e. same profile but $100\%$ slower} associated to the frequencies $\varphi=[ 0.05,\: 0.15,\: 0.3,\: 0.4, \:0.55,\: 0.7,\: 0.9,\: 1.1, \:1.35, \:1.75, \:2.2 ]$.
\end{enumerate}
in summary, such a definition of FRF \textcolor{black}{discussed here} differs slightly from a usual discrete transfer function because it is defined on the \textit{custom} frequencies $\varphi$.

The presented definition of FRF is important in posture control literature as it is used in \cite{peterka2002sensorimotor} and several studies that took inspiration from \cite{peterka2002sensorimotor}. For example \cite{asslander2020reductions,ketterer2024effects,missen2024velocity,asslander2015analysis,goodworth2021postural}. The choice of averaging groups of frequencies was initially motivated by plotting the FRF on a logarithmic scale \cite{goodworth2021postural}, where the higher frequency bands contain many more data points than lower frequency bands and these higher frequency points had lower signal to noise. Averaging more adjacent high-frequency points together compared to the averaging of adjacent low-frequency points makes the final set of FRFs approximately equally spaced on a logarithmic frequency scale and had similar confidence intervals \cite{goodworth2021postural, goodworth2009contribution,otnes1972digital}. This implies that in analyzing and comparing the FRFs, higher frequencies are considered less important within the experiment's scope. This is a reasonable assumption in many cases, for example, in experiments with the subject standing on a tilting support surface, where the dynamics of the body have a low-pass characteristic. In some cases, it is preferable to avoid averaging. For example in some works the FRFs are computed and plotted on equally spaced frequencies for reasons due to the purpose of the study: in \cite{lippi2018prediction} to analyze the posture control behavior of a humanoid that can exhibit oscillations due to delay; in \cite{lippi2020real} to evaluate the effect of the tracking system; in \cite{alexandrov2017human} to study the theoretical response of the posture control of a humanoid robot.
Note that the method proposed in the present work does not require the groups of frequencies to be averaged and can be applied to any kind of transfer function estimation as long as it can be {transformed into a real function in the time domain}, as discussed in the next section.

\subsection{The Bootstrap Method}
\begin{figure}[tb!]
\centering
\includegraphics[width=1.00\textwidth]{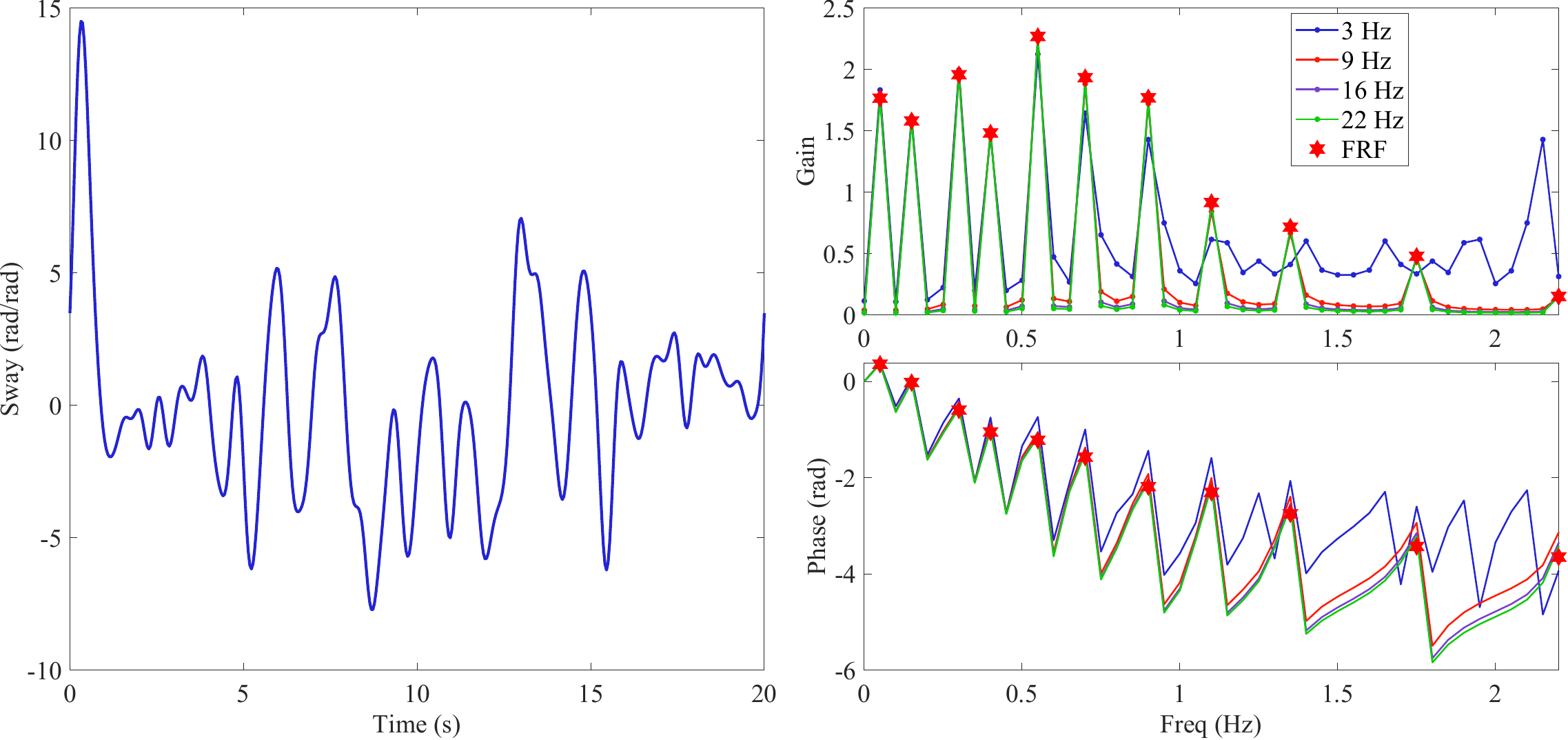}
\caption{The PIR on the left, the FRF (red stars), and the DFT of the PIR (colored lines). The PIR is computed with eq. \ref{pseudo}. Different sample times are tested to reconstruct the FRF through a DFT. The period of the PIR is defined as the inverse of the greatest common divisor of the frequencies in $\varphi$; the sample time used in the examples is set to ten times the highest frequency in $\varphi$, i.e., 22 Hz. Notice how the gain tends to converge to zero between the peaks}
\label{Pseudopulse}
\end{figure}

The prediction and confidence bands are computed on the {basis} of $n$ FRFs. The $i$th FRF is denoted as $H_{i,f}$.
The experimentally observed FRFs are considered to be perturbations of a true FRF $\hat{H}$ that can be seen as a discrete Fourier transform of a time domain function. frequencies, retains the property of being the Fourier transform of a real signal. The proof is straightforward: given the empirical transfer function $\Phi(f_k)=G(f_k)_{xy}/G(f_k)_{yy}$, the FRF with averaged values over frequency bands is defined as $H(\varphi_i)=\sum_{k \in \beta_i}\Phi(f_k)$ where $\beta_i$ is the set of indexes defining the frequency band averaged to compute the $i^{th}$ element of $H$. As $\Phi(f_k)$ is by definition the {Fourier} transform of a real value and both $\Phi$ and $H$ are represented on positive frequencies the averaging operation above preserves the property of even symmetry and hence also $H$ has a real {inverse} DFT. By analogy to the Fourier transform of transfer functions defined on a continuous range of frequencies that represents the impulse response function of a linear system, the {inverse} Fourier transform of the discrete FRF can be seen as a response to an impulse, {the PIR mentioned in the introduction. Such function, is estimated with a continuous input stimulus and being a periodic function, is not the actual impulse response of the system, for this reason is defined \textit{pseudo
}-impulse-response.} The important property of the PIR is that it is a time domain function that contains all the information included in the FRF. The PIR can be computed as
\begin{equation}
\label{pseudo}
x_i(t_j)= \sum^{M}_{k=1} \Re(H_{i,k}) \cos(2\pi F_{k} t_j) + \Im(H_{i,k})\sin(2\pi F_{k} t_j)
\end{equation} 
In the present work, $M=11$ following the structure of the FRF presented in \cite{lippi2023human,lippi2020human}. {In} the original formulation \cite{peterka2002sensorimotor}, the FRF had 22 components. The frequency $F_k$ is $k$th component of the vector $\varphi$. By eq. \ref{pseudo} the $i$th FRF, $H_{i}=[H_{i,1}, H_{i,2} ... H_{i,M}]$ is transformed into $x_i(t_j)$ that can be seen as a vector of coefficients defining the FRF as the time $t_j$ is defined on a discrete range \textcolor{black}{(to address this, the variable $t_j$ has the index $j$)}. Specifically, the period of the PIR is defined as the inverse of the greatest common divisor of the frequencies in $\varphi$. Figure \ref{Pseudopulse} shows how the FRF can be reconstructed by applying the DFT to the PIR with different numbers of time samples. \textcolor{black}{The sample time in the examples is chosen to have ten times the highest frequency in $\varphi$ to provide a sufficiently accurate reconstruction}. The PIR is equivalent to the FRF as it can be transformed back to the frequency domain by a DFT, leading to a complex function that is equal to the FRF {at} the frequencies in $\varphi$ and zero elsewhere as shown in Fig.\ref{Pseudopulse}.
The statistical analysis of the PIRs allows an analysis of the FRF distribution that combines all the frequencies together. In \cite{goodworth2009contribution}, an impulse response function is computed using the PRTS stimulus through a scaled cross correlation between the PRTS velocity waveform and the response waveform \cite{davies1970system}. This was done to produce a time domain representation of the response equivalent to the FRF, making it easier to visualize some features, such as time delay, that are less evident in the frequency domain. The PIR is similar but slightly different because it is not computed directly on the time domain samples but on the FRF that includes the averaging between stimulus cycles (and can be the average between different trials by the same subject) and the averaging over frequency ranges. This is because the PIR is intended to be an instrument to perform statistical tests specifically on the FRFs as represented in the specific study.
Given {a} set of PIRs, the mean and the standard deviation can be estimated at each sample time to describe the distribution as follows:
\begin{align}
\label{variability}
\hat{x}(t_j)= \frac{1}{N} \sum^{i=1}_{N} x_i(t_j) \\
\hat{\sigma}_x(t_j) = \sqrt{\frac{1}{N-1}\sum^{i=1}_{N} \left|x_i(t_j)-\hat{x}(t_j) \right|^2} \label{sigmahat}
\end{align}
Once the quantities above are defined (and computed), the prediction and the confidence bands can be constructed.
\textbf{The prediction band} can be defined for a new draw from the FRF distribution $H_{N+1}$, and hence for the respective PIR $x_{N+1}(t_j)$. {Following the methods presented in \cite{lenhoff1999bootstrap}}, given the desired confidence level $\alpha\%$, the constant $C_p$ is defined to obtain the probability
\begin{equation}
P\left[ \max\limits_{t_j} \left( \frac{|x_{N+1}(t_j) - \hat{x}(t_j)|}{\hat{\sigma}_{x}(t_j)} \right) \leq C_p\right] = \frac{\alpha}{100}
\label{confidenceeq}
\end{equation}
\textcolor{black} {where $\max\limits_{t_j}$ selects the maximum value assumed by the expression in parenthesis over the range of values $t_j$.} The $\alpha\%$ prediction band for a new \textcolor{black}{PIR} can be computed as
\begin{equation}
\hat{x}(t_j) \pm C_p \cdot \hat{\sigma}_{x}(t_j)
\end{equation}
\textbf{The confidence band} can be defined similarly choosing the $C_c$ in order to obtain
\begin{equation}
P\left[ \max\limits_{t_j} \left( \frac{|x(t_j) - \hat{x}(t_j)|}{\hat{\sigma}_{x}(t_j)} \right) \leq C_c\right] = \frac{\alpha}{100}
\label{predictioneq}
\end{equation}
The $\alpha\%$ confidence band for $\hat{x}(t_j)$ is then
\begin{equation}
\label{confidenceconstant}
\hat{x}(t_j) \pm C_c \cdot \hat{\sigma}_{x}(t_j)
\end{equation}
\textbf{The bootstrap} is used to determine $C_c$ and $C_p$. Approximated versions of the probabilities in eq.\ref{confidenceeq} and \ref{predictioneq} are obtained using empirical distributions produced by resampling the sample set. The constants $C_c$ and $C_p$ are set so that the approximated probability is as close as possible to the desired confidence $\alpha\%$. Specifically eq. \ref{confidenceeq} has the following bootstrap approximation \cite{lenhoff1999bootstrap}:
\begin{equation}
\frac{1}{B}\sum\limits_{b=1}^{B}\left[\frac{1}{n}\sum\limits_{i=1}^{n}I\left(\max\limits_{t_j} \left( \frac{|x_i(t_j) - \hat{x}^b(t_j)|}{\hat{\sigma}^b_{x}(t_j)} \right) \leq C_p\right)\right]
\label{bootstrapped}
\end{equation}
where I(E) is 1 or 0 according to \textcolor{black}{whether} condition E is or is not verified, respectively. Eq. \ref{bootstrapped} is the average, over the $B$ bootstrap replications, of the proportion of the original data curves whose maximum standardized deviation from the bootstrap mean is less than or equal to $C_p$. The superscript $^b$ \textcolor{black}{indicates} that the quantity is computed on the basis of the resampled set. Notice that $\hat{\sigma}^b_{x}(t_j)$ is based on the resample set at every iteration of the bootstrap as recommended in \cite{hall1991two}. Such ``pivotization'' also allows for null hypothesis testing without having to simulate the distribution produced by the null hypothesis \cite{davison2003recent}, as will be shown in the examples. The number $B$ is set to be relatively large with a trade-off between the accuracy and the computational time. Different indications about the required $B$ are discussed in \cite{zoubir1998bootstrap}. As a practical rule, it should be considered that the number of resamplings sets a limit on the possible resolution of $\alpha\%$, e.g., if there are $3$ samples and one asks which portion of them are above a given threshold, the only possible outcomes are $0$, $1/3$, $2/3$ and $1$. Furthermore, there is no advantage in choosing $B$ larger than the number of combinations with repetitions of the $N$, i.e. $N^N/N!$. However, this is relevant just for very small (possibly too small) data sets. In previous posture control experiments \cite{lippi2023human} and \cite{lippi2020body}, the components were tested independently, and several resampled sets in the order of $B=10^4$ produced acceptable results with groups of $7$ and $36$ subjects, respectively. The specific bootstrap included a nested resampling $B_{NEST}=200$ used to estimate the variance. In the present work, such nested bootstrap is not used as the variability is quantified by $\hat{\sigma}^b_{x}(t_j)$ defined in eq. \ref{sigmahat}.
Similarly eq. \ref{predictioneq} can be approximated as
\begin{equation}
\frac{1}{B}\sum\limits_{b=1}^{B}\left[I \left( \max\limits_{t_j} \left( \frac{|\hat{x}(t_j) - \hat{x}^b(t_j)|}{\hat{\sigma}^b_{x}(t_j)} \right) \right) \leq C_c\right]
\label{confidencebootstrap}
\end{equation}

\subsection{Sample Sets and Tests}
\label{section:tests}
\begin{figure}[tb!]
\centering
\includegraphics[width=1.00\textwidth]{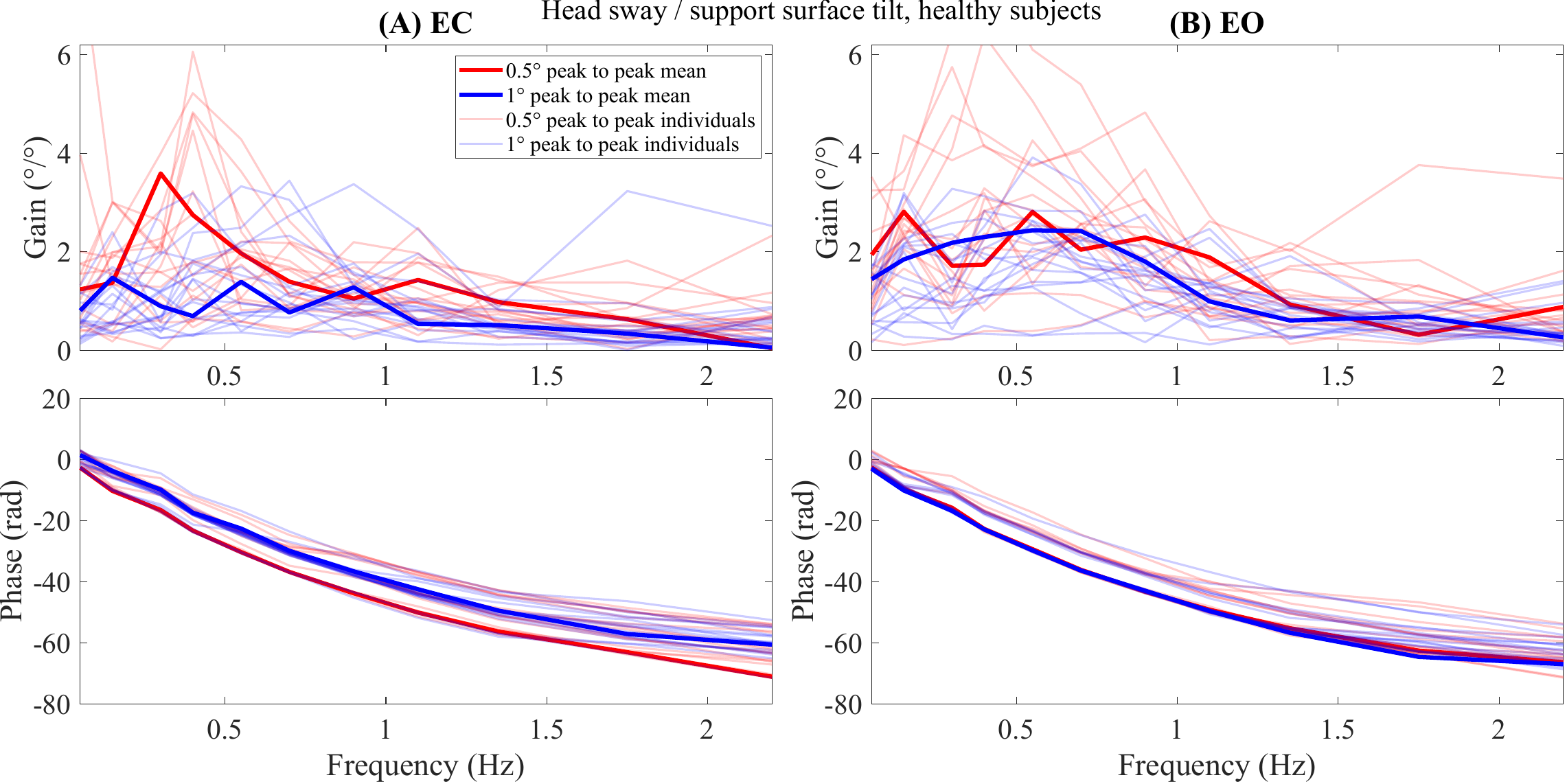}
\caption{Test set to demonstrate confidence bands and their use. The FRFs of head sway induced by support surface tilt from \cite{icinco23} (with permission) are shown for the condition eyes closed, (A) EC, and eyes open, (B) EO. The red color indicates the FRFs obtained with the PRTS stimulus with \textit{peak to peak} amplitude of $0.5^{\circ}$, the blue with $1^{\circ}$. The thicker, darker lines represent the mean of the two groups of samples of the same color.}
\label{fig:StatExample}
\end{figure}

\begin{figure}[tb!]
\centering
\includegraphics[width=1.00\textwidth]{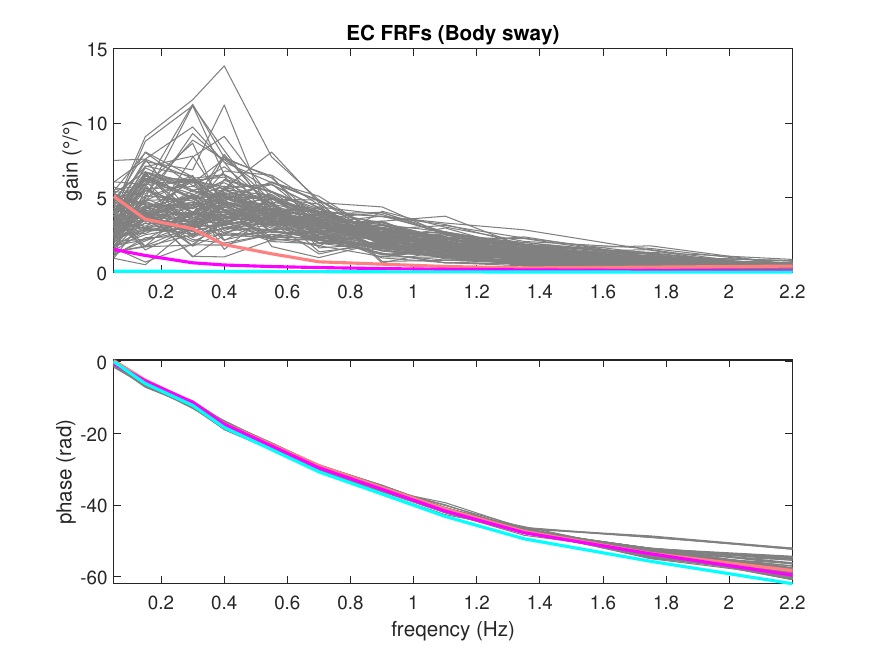}
\caption{Data prediction (body sway responses to support surface tilt from \cite{lippi2023human,lippi2020human,robovis21}, with permission) in gray, plotted together with three responses from simulations \cite{lippi2020human}. The three simulated responses implement three different control systems: Disturbance Estimation and Compensation (DEC) \cite{lippi2017human} in coral, independent channel (IC) \cite{peterka2002sensorimotor} in magenta, and Eigenmovements (EM) \cite{alexandrov2017human} in cyan.}
\label{fig:SampleDataNeck}
\end{figure}

In order to test the approach, data sets obtained from human experiments with healthy subjects are used. Specifically:
\begin{itemize}
\item {\textbf{Simulated low pass filter data} produced as an example of the use of the residuals to visualize the difference between a tested sample and the confidence intervals (see Section \ref{section:residuals}). The responses are produced to have FRFs with low pass profiles with different cut-off frequencies. Having such difference known a priori allows one to test how it is reflected in the visualization of the results in frequency domain. 50 Samples have been generated for each group, adding a Gaussian noise with $\sigma=0.5$ to predefined amplitude signals with magnitude 1 up to  $0.4 Hz$ and $0.75 Hz$, respectively, and imposing minimal phase on each sample.}
\item \textcolor{black}{the \textbf{confidence bands} are tested on FRFs from head sway induced by support surface tilt in 18 healthy subjects\cite{icinco23}, shown in Fig.\ref{fig:StatExample} . For each subject, a trial is performed with eyes open and closed.  \textcolor{black}{A paired test is performed: the trials are matched by subject, and the confidence bands are computed on the difference between the FRFs (and hence the PIRs)}. According to the null hypothesis that the two groups of trials have the same distribution, the average is expected to be zero. Comparing the horizontal line, $PIR=0$, versus the bands provides a test for rejecting the null hypothesis with $\alpha\%$ confidence. The test is done with peak-to-peak stimulus amplitude $1^{\circ}$ and $0.5^{\circ}$.}
\item the \textbf{prediction bands} are tested on 113 FRFs from body (center of mass) sway responses to support surface tilt \cite{lippi2023human,lippi2020human,robovis21}. Three FRFs from simulated responses from \cite{lippi2020human} (with permission) have been compared to the prediction bands to evaluate their  \textcolor{black}{similarity to experimental results from humans}. Leave-one-out cross-validation has been performed to check the validity of the bands obtained with the bootstrap. The FRFs of the human data set and the three simulated responses are shown in Fig. \ref{fig:SampleDataNeck}.
\end{itemize}

\subsection{Visualizing the Results in Frequency Domain}
\label{section:residuals}
\begin{figure}[htbp]
	\centering
		\includegraphics[width=1\textwidth]{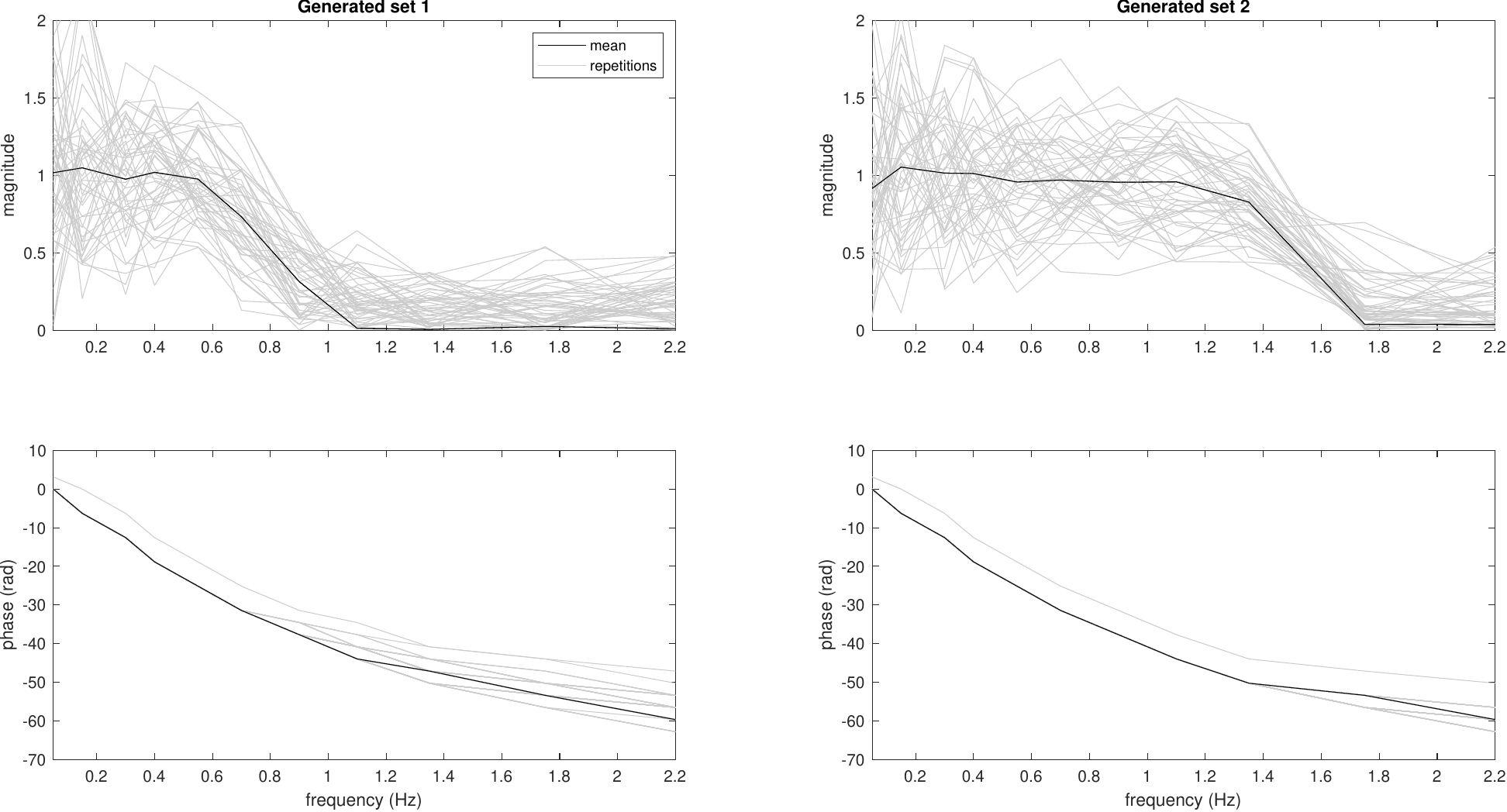}
	\caption{FRFs with low pass profiles with different cut-off frequencies. 50 Samples have been generated for each group, adding a Gaussian noise with $\sigma=0.5$ to predefined amplitude signals with magnitude 1 up to  $0.4 Hz$ and $0.75 Hz$, respectively, and imposing minimal phase on each sample.}
	\label{resgen_DATA}
\end{figure}

Using the bands defined on the PIRs for tests provides a result visualized in the time domain without immediate interpretations. Mapping the result in the frequency domain allows for considerations about how the different distribution sets (subjects groups or conditions) differ in specific frequency ranges (e.g., in \cite{joseph2014contribution} where bands like \textit{low frequencies}, \textit{middle frequencies} and \textsl{high frequencies} are defined manually). A post-hoc test can be performed on single frequencies as proposed in \cite{asslander2014sensory}. A further possibility can be plotting the DFT of the average PIR after a test with confidence bands or the DFT of the difference between the average PIR of the reference set and the PIR of the tested sample tested with prediction bands.
Here a visualization method based on the use of the confidence bands is proposed. The idea is to plot the residuals between the PIR and the confidence bands.
 A residual can be defined as the difference between a PIR $x_i(t_j)$ that exceeds the bands:
\begin{equation}
r(t_j)=\left\{\begin{array}{cl}
x_i(t_j) -(\hat{x}(t_j) + C_c \cdot \hat{\sigma}_{x}(t_j)),&x_i(t_j) \geq \hat{x}(t_j) + C_c \cdot \hat{\sigma}_{x}(t_j)\\
0,& |\hat{x}(t_j) - x_i(t_j)| \leq C_c \cdot \hat{\sigma}_{x}(t_j)\\
x_i(t_j) -(\hat{x}(t_j) - C_c \cdot \hat{\sigma}_{x}(t_j)),&x_i(t_j) \leq \hat{x}(t_j) - C_c \cdot \hat{\sigma}_{x}(t_j)\end{array}\right.
\label{eq:residual}
\end{equation}
The DFT of $r(t_j)$, specifically on the frequencies $\varphi$, provides a frequency domain plot of how the sample exceeded the bands. While the complex function does not have an easy interpretation, its gain or power spectrum can be easily used to visualize how different frequency components contribute to the difference between the sample and the confidence bands. Artificial data have been designed to 
provide an example (Fig. \ref{resgen_DATA}) The residuals have the advantage of being different from zero only if the sample exceeds the confidence interval and represents just the part of the signal that determines the test outcome. On the other hand, the definition of the residuals in eq. \ref{eq:residual} is a nonlinear operation applied on the signal that produces a distortion reflected in the frequency domain and introduces components {at} frequencies where the original signal had no power. The difference between the FRFs' means, being a linear operation, has a more straightforward interpretation in that it is equivalent to the DFT of the differences between the PIRs. {In the Results section the two methods of visualization, residuals and difference of the means are compared in Fig. \ref{resgen_RESULT} and Fig. \ref{fig:SampleStatConfSingle}.}

\section{Results}
\label{section:results}
The sample tests described in \S \ref{section:tests} show the application of the method. Each test reported here has been performed with $B=10000$. Tests with smaller and larger values of $B$ (i.e., $1000$ and $50000$) produced similar results to the ones reported in the following.

\begin{figure}[tb!]
\centering
\includegraphics[width=1.00\textwidth]{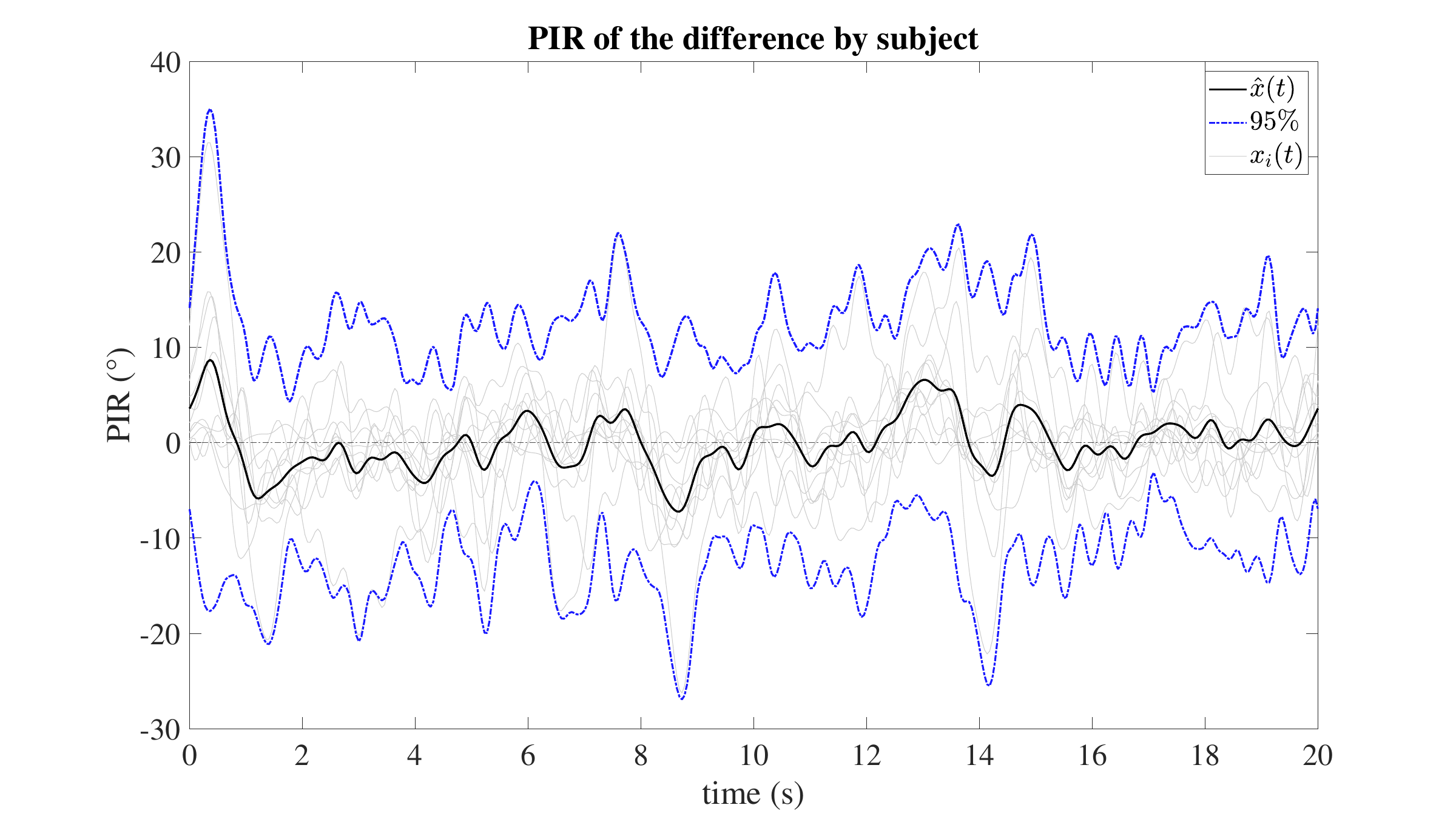}
\caption{Example of confidence band and its use. The PIRs are computed on the difference between the EO and EC FRFs for peak to peak stimulus amplitude $0.5^{\circ}$ shown in Fig. \ref{fig:StatExample}, as the data are matched (i.e., same subject, different condition). The $95\%$ confidence interval of the mean of the PIRs is compared with zero ($H_0$ : the average is zero). Null values (zero) {at some} time points are entirely inside the confidence bands, meaning that the two conditions did not produce a significantly different response with $p<.05$.}
\label{fig:StatExampleResult}
\end{figure}

\begin{figure}[tb!]
	\centering
		\includegraphics[width=1.00\textwidth]{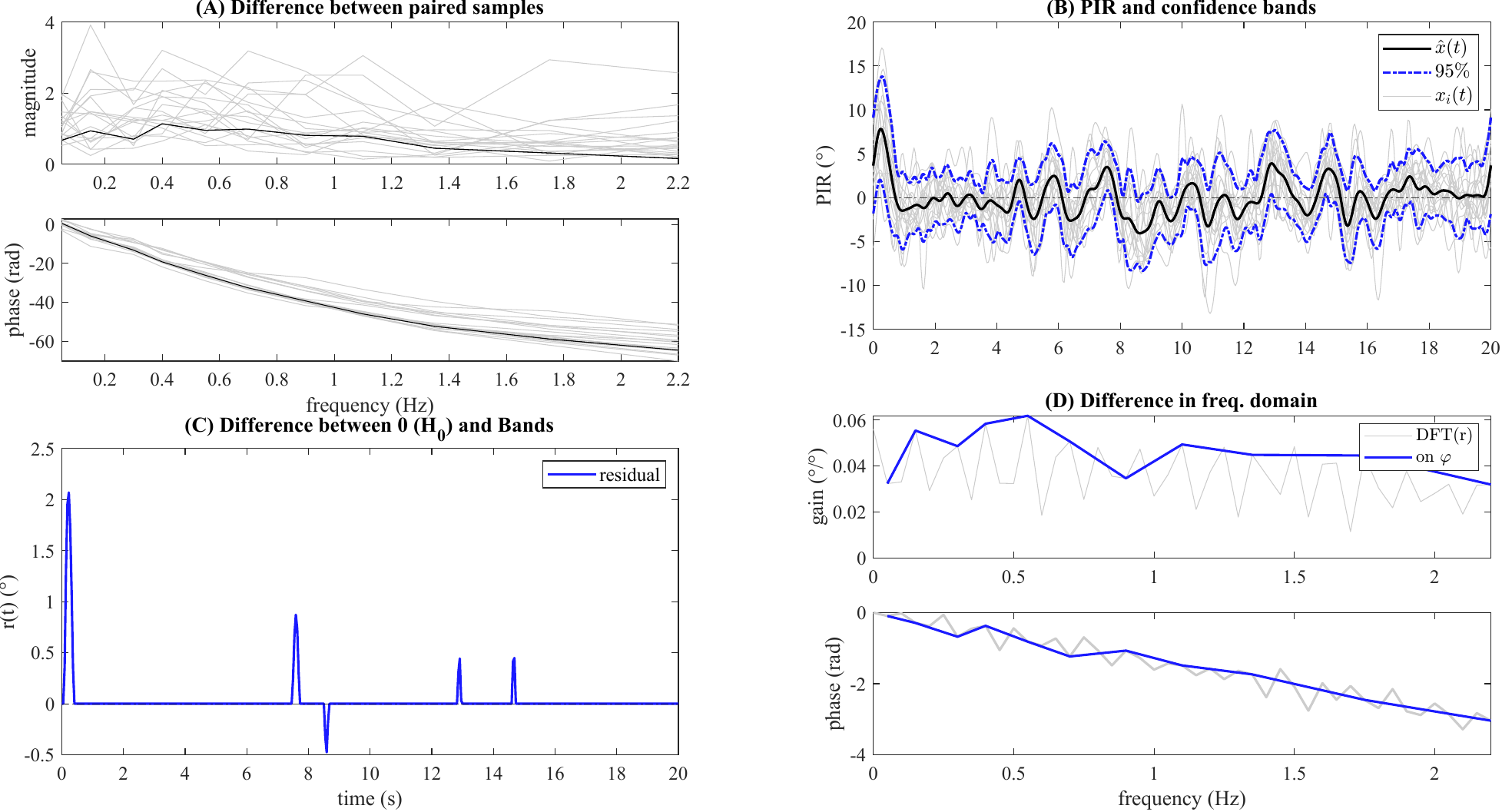}
	\caption{Null hypothesis rejection for paired samples based on confidence bands: A set of 18 FRFs, each one computed as differences between two FRFs produced by the same subject in response to the same stimulus under different conditions (EC vs EO). The hypothesis that such a difference has a null average (i.e., no significant difference induced by the different conditions) is rejected. The data used are the head sway response to small stimuli from \cite{lippi2020human,robovis21}, a subset of the FRFs shown in Fig. \ref{fig:SampleDataNeck}, specifically the ones with peak-to-peak amplitude $1^{\circ}$.(A) Difference between the samples (paired). (B) The PIR of the head sways in response to FRFs. The $95\%$ confidence interval of the mean of the PIRs is compared with zero ($H_0$ : the average is zero). Null values (zero) across all time points are outside the confidence bands, meaning that the effect of the stimulus is significantly different from zero ($p<.05$). (C) represents the difference between 0 and the boundaries functions, eq. \ref{confidenceconstant}, defining the confidence bands. Such residual function $r(t_j)$ can be transformed back to the frequency domain and plotted over the frequencies $\varphi$ as shown in (D). The gain {provides an idea of the different} contribution of the FRFs' components to the significant difference between the conditions.}
	\label{fig:SampleStatConfSingle}
\end{figure}

\textbf{The confidence band} is used on matched trials. The same healthy subject is tested under two conditions: eyes open and closed. The measured variable is the sway of the head. The bands and their comparison with the null hypothesis (i.e., no difference, leading to a null average) are shown in Fig. \ref{fig:StatExampleResult} \textcolor{black}{for peak-to-peak amplitude $0.5^{\circ}$ and in Fig.\ref{fig:SampleStatConfSingle} for peak-to-peak amplitude $1^{\circ}$. The null hypothesis can not be rejected with $95\%$ confidence in the former case, while it is rejected in the latter. In Fig.\ref{fig:SampleStatConfSingle} The residual $r(t_j)$ from eq. \ref{eq:residual} is plotted together with its DFT, showing that most difference between the two conditions is expressed in lower frequencies (i.e., where the gain is larger).}

\begin{figure}[tb!]
\centering
\includegraphics[width=0.80\textwidth]{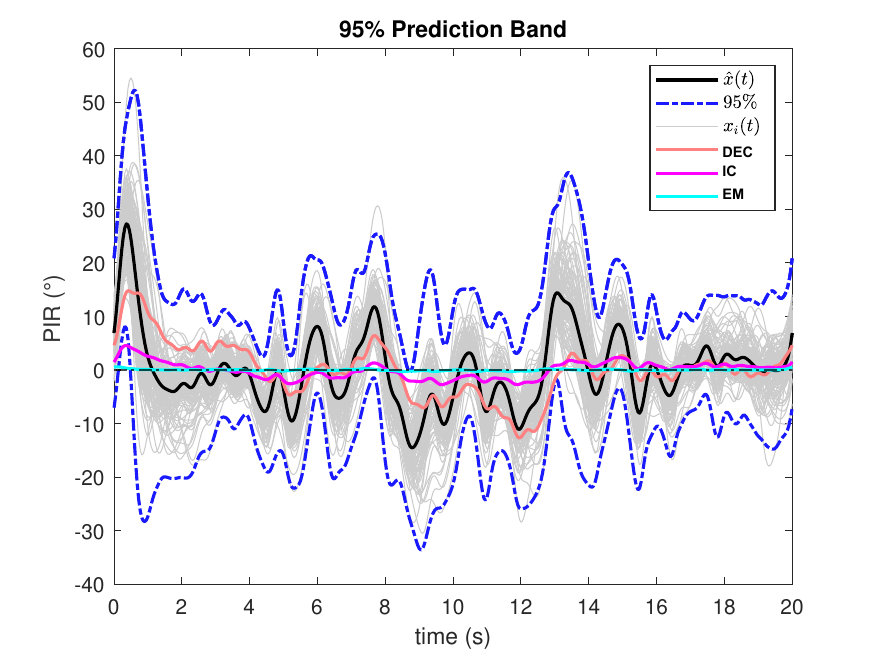}
\caption{Prediction bands with the three sample FRFs from \cite{robovis21}. The light grey lines represent the PIR of all the FRFs in the dataset. The black line represents the average PIR, and the blue dotted lines represent the boundaries of the $95\%$ prediction bands. The PIRs of the three simulated responses are superimposed for comparison.}
\label{fig:TimeDomainPrediction}
\end{figure}

\textbf{The prediction bands} are computed on the data shown in Fig. \ref{fig:SampleDataNeck}. In Fig. \ref{fig:TimeDomainPrediction}, the prediction bands are shown together with the PIR of the three simulated sway responses. The bands are shown in Fig. \ref{fig:TimeDomainPrediction} together with the PIR of the three simulated tests. Only one of the three tested samples was within the $95\%$ confidence prediction band.

\textbf{A cross-validation} has been performed to evaluate the prediction bands. Specifically, a leave-one-out cross-validation was used. The prediction bands have been computed 113 times, each using 112 samples and leaving a sample out for testing. The coverage is estimated as the percentage of test samples that are inside the bands. The value obtained is $93.8\%$.

\begin{figure}[htbp]
	\centering
		\includegraphics[width=1\textwidth]{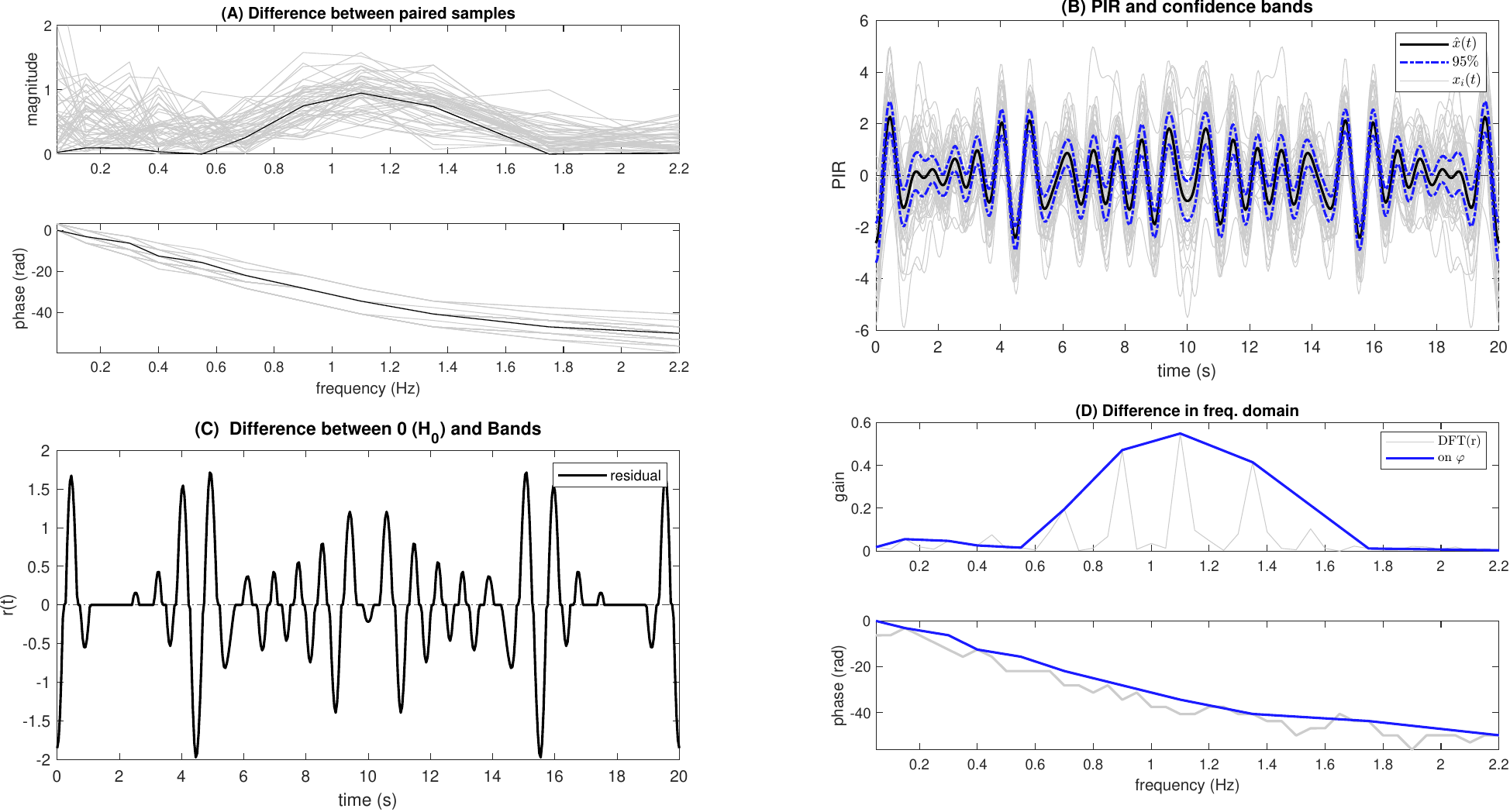}
	\caption{An example of the use of the residuals to visualize the difference between a tested sample and the confidence intervals.Data shown in Fig. \ref{resgen_DATA} were used. (A) Difference between the samples (paired). (B) PIR and confidence bands in comparison with H0 ($x(t)=0$). (C) Residuals in the time domain. (D) DFT of the residuals.}
	\label{resgen_RESULT}
\end{figure}

 {
\subsection{Visualization in Frequency Domain} An example is provided using artificial data designed to be different in a way that is known a priori, i.e. two sets {of} low pass FRFs with a different cut-off frequency (Fig. \ref{resgen_DATA}). Both the plot of the difference of the FRF groups means (Fig. \ref{resgen_RESULT}A) and the DFT of the residuals (Fig. \ref{resgen_RESULT}D) show a peak in the range of frequency where the two groups are different ($1$ to $1.5 Hz$). The $r(t_j)$, associated with the confidence bands for a test that rejects the null hypothesis in the results, is shown in Fig. \ref{fig:SampleStatConfSingle} (B). The DFT of $r(t_j)$ is shown in Fig. \ref{fig:SampleStatConfSingle} (C). }
\section{Discussion and future work}
The proposed tests showed that implementing confidence and prediction bands in the time domain using the PIRs allows statistical tests to answer the questions typically associated with posture control studies. Specifically, the PIR considers all the frequencies together and, being defined in the real domain, does not require imposing arbitrary choices to perform statistics on real and complex values together. Additionally studying the DFT of the residual $r(t_j)$ (eq. \eqref{eq:residual}) can be used to plot and discuss the result in the frequency domain.

The prediction bands can be used to compare single trials to a reference distribution. \textcolor{black}{In the {results} presented in Fig. \ref{fig:TimeDomainPrediction}, prediction bands were used to assess the the similarity of human results to three stimulated results that represent different robotics balance control schemes \cite{robovis21}}. The same test can be applied to exoskeletons and other assistive devices or as a diagnosis tool. All those cases require a choice of the $\alpha\%$ confidence level that, here set to $95\%$ as the value traditionally accepted as significant. Future work may include developing a function that estimates the probability of a particular FRF, and hence the relative PIR, to be generated by a specific distribution of FRFs.\textcolor{black}{ The use of different input profiles besides the PRTS can be considered in the future, e.g., raised cosine\cite{10.3389/fnbot.2018.00021}, sum of sinusoids \cite{MAKI19931181} or impulsive stimuli (robot example in \cite{monteleone2023method}). Also, another kind of pseudorandom stimuli can be used as in \cite{10.3389/fnhum.2024.1471132}, although how to treat multidimensional stimuli is still an open issue. The frequency domain may not be the usual or best representation for all of those stimuli. However, in some studies, it may be interesting to use the proposed analysis on the results produced by different stimuli on the same subject (or humanoid) to compare the results. In most cases, an opportune definition of the frequencies $\varphi$ would be reasonably enough to adapt the method.}

Two methods, residuals and differences of the means, have been proposed to visualize the differences between samples tested using the confidence intervals in the frequency domain. Once the confidence intervals are exceeded, whether to use residuals, mean differences, or a component-by-component bootstrap test depends on the specific experimental context. This choice should be made carefully considering the limitations and interpretive risks associated with post-hoc analysis.

As many posture control experiments consider more than one degree of freedom, in the future, the approach may be extended to allow for multivariate analysis, i.e., considering more joint responses, for example, trunk sway and leg sway, at once.

\section{The Code}
\begin{figure}[tb!]
\centering
\includegraphics[width=0.80\textwidth]{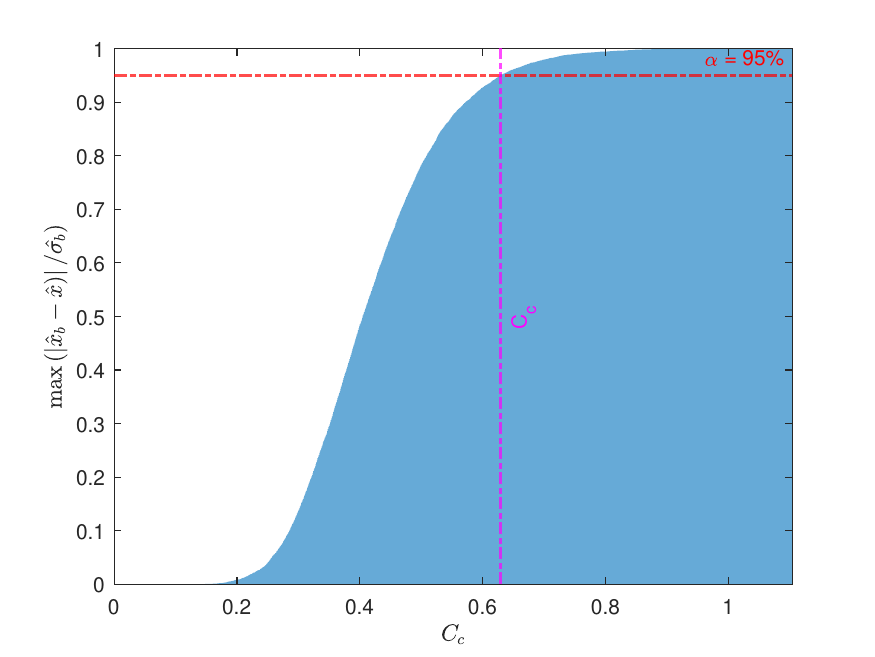}
\caption{Cumulative Histogram for the terms of the sum in eq. \ref{confidencebootstrap} and the consequent choice of $C_c$ set to make eq. \ref{confidencebootstrap} empirically match the desired confidence $\alpha$. In this example, $\alpha = 95\%$ as presented in the Results section, \S \ref{section:results}.}
\label{fig:HistogramConf}
\end{figure}

The code provides two functions to compute the confidence and the prediction bands.
The functions are written in Matlab 2019 \cite{MATLAB:2019b}, It is reasonably compatible with previous versions from 2014a where the function \texttt{histogram()} used in the functions was defined for the first time.
\textbf{The confidence band} is computed with the functions
{\footnotesize
\\
\\
\texttt{[avg,sigma,band,Cc,chist,values]=FRF\_ConfidenceBand(FRFs,phi,sample\_time,alpha,B)}
\\
\\}
where \texttt{avg} is average PIR, \texttt{sigma} is the measure of variation $\hat{\sigma}_x(t_j)$, \texttt{band} is a two row matrix with the boundaries of the bands $C_c\cdot \hat{\sigma}_x(t_j)$. \texttt{Cc} is the constant $C_c$ obtained by the bootstrap. The constant $C_c$ is computed by ordering the values in the argument of the sum in eq. \ref{bootstrapped} produced by each bootstrap repetition and computing a cumulative histogram on which the constant is chosen to obtain the desired confidence $\alpha$ as shown in Fig. \ref{fig:HistogramConf}. \texttt{FRFs} is a matrix where each row represents a FRF of the set, \texttt{phi} is the vector of frequencies $\varphi$ and \texttt{sample\_time} is the sample time of the PIRs, set by default to ten times the highest frequency in $\varphi$. \texttt{chist} is a vector representing the cumulative histogram for the values returned in \texttt{values}.

\textbf{The prediction band} is computed similarly to the confidence band with the function
{\footnotesize
\\
\\
\texttt{[avg,sigma,band,Cp,chist,values]=FRF\_PredictionBand(FRFs,phi,sample\_time,alpha,B)}
\\
\\}
Specifically, in this function, the vectors \texttt{values} and \texttt{chist} returned are the $n \times B$ values computed inside the nested sums in eq. \ref{bootstrapped}.

\textbf{Computing the PIR} is performed by the service function 
\\
\\
\texttt{[x,t]=FRF\_pseudoimpulse(FRF,Freq,sf)}
\\
\\
Where \texttt{FRF} is a single FRF, \texttt{Freq} is the vector $\varphi$ of frequencies and \texttt{sf} is the sample frequency. The function returns the PIR \texttt{x} and the time vector \texttt{t}.
 
\textbf{Plotting the bands} is possible through the dedicated function
\\
\\
\texttt{h=FRF\_Plotbands(avg,band)} \\ \\
that provides a graph of the bands like the ones presented in Fig. \ref{fig:StatExampleResult} (A). The return value \texttt{h} is the handle to the generated plot.

\textbf{The source code is available} in the repository: https://github.com/mcuf-idim/FRF-statistics While the present paper was under review some additional functions have been added to the library, they are presented in \cite{newFRFpaper}.
\bibliography{bibliography} 

\end{document}